\documentclass{appolb}
\usepackage{graphicx}

\usepackage{epsfig}
\usepackage{float}
\usepackage{amssymb}
\usepackage{hyperref}
\usepackage{amsmath}
\usepackage{caption}
\usepackage{subcaption}
\usepackage{setspace}

\begin{document}
	\title{Probing flow fluctuations through factorization breaking of harmonic flows in heavy-ion collision%
		\thanks{Presented at Quark Matter 2022 : XXIXth International Conference on Ultra-relativistic Nucleus-Nucleus Collisions }%
	}
	\author{Piotr Bożek, Rupam Samanta
		\address{AGH University of Science and Technology, Faculty of Physics and
			Applied Computer Science, aleja Mickiewicza 30, 30-059 Cracow, Poland}
		\\[3mm]
	}
	\maketitle
	\begin{abstract}
		
		We study factorization-breaking coefficients between the momentum dependent and momentum averaged flow vectors to probe  flow fluctuations caused by  initial-state fluctuations in heavy-ion collision. The coefficients for the flow vector \textit{squared} and flow magnitude \textit{squared} could be used for the extraction of flow angle decorrelations. We compare our model results with preliminary experimental data. We also present the predictions for the momentum dependent correlation between mixed flow harmonics.
	\end{abstract}
	
\section{Introduction}
	  The dense matter created in the interaction region of a heavy-ion collision, expands rapidly. Due to the asymmetry of the initial source distribution, the collective expansion leads to an azimuthal momentum anisotropy of the spectra of  particles,
	       \begin{equation} 
		      \frac{dN}{dp d\phi} = \frac{dN}{2 \pi dp} (1+ 2 \sum_{n = 1}^{\infty}V_n(p)e^{i  n  \Psi_n})
		      \label{eq: flowharmonic}
	       \end{equation}
	   where the Fourier coefficients $V_n(p)'s$ are called the harmonic flow coefficients. 
	   
	   The initial source fluctuates event-by-event (ebe), leading to the ebe fluctuation of the harmonic flow coefficients in the final state \cite{PHOBOS:2007vdf,Alver:2010gr}. These flow fluctuations can be probed by studying the correlation coefficient between the harmonic flow vectors in two different transverse momentum bins \cite{Gardim:2012im}, the  \textit{factorization-breaking coefficients}. The fluctuations result in the decorrelation (deviation of the correlation coefficients from 1) between these flow vectors, which manifests the combined effect of flow magnitude decorrelation as well as the flow angle decorrelation\cite{Jia:2014ysa,Bozek:2017qir}. These two effects can be separated by constructing four particle correlators between two transverse momentum bins. However, to avoid the experimental difficulty due to the low statistics at higher momenta, the coefficients are constructed keeping two particles fixed at transverse momentum bins (differential flow) and the other two are momentum averaged (integrated flow)\cite{NielsenIS2021}. We compare our model results with the preliminary results by ALICE collaboration. We also study the similar factorization coefficients between the mixed flow harmonics e.g. $V_2^2$ - $V_4(p)$ or $V_2 V_3$ - $V_5(p)$ , which serve  as measures of the non-linear hydrodynamic response of the medium.
	   
\section{Factorization Breaking for harmonic flow vector, magnitude and angle}
The complex flow vector $V_n(p)=v_n(p)  e^{i n \Psi_n(p)}$ in Eq. \ref{eq: flowharmonic} denotes the harmonic flow vector of order $n$ for particles emitted at the transverse momentum $p$; $v_n$ and $\Psi_n$ are the corresponding flow magnitude and angle. In general , one can construct the four-particle correlators, measuring the factorization breaking coefficients between two flow vectors \textit{squared} at two different transverse momentum bins as \cite{Gardim:2012im}
   \begin{equation}
	r_{n;2}(p_1,p_2)=\frac{\langle V_n^2(p_1) V_n^{\star\ }(p_2)^{2} \rangle}{\sqrt{\langle v_n^4(p_1)  \rangle \langle v_n^{ 4}(p_2) \rangle}}
   \end{equation}
and similarly for the flow magnitude \textit{squared} , the factorization-breaking coefficient can be constructed as \cite{Bozek:2018nne},
\begin{equation}
		r_{n}^{v_n^2}(p_1,p_2)=\frac{\langle |V_n(p_1)|^2 |V_n(p_2)|^2 \rangle}{\sqrt{\langle
				v_n^4(p_1)  \rangle \langle v_n^{ 4}(p_2) \rangle}} = \frac{\langle v_n^2(p_1) v_n^{ 2}(p_2) \rangle}{\sqrt{\langle v_n^4(p_1)  \rangle \langle v_n^{ 4}(p_2) \rangle}}
\end{equation}
However, in practice, the above formulas are difficult to use in experiment due to the low statistics in higher momentum bins.

\begin{figure}
	\centering
	\begin{subfigure}{.5\textwidth}
		\centering
		\includegraphics[height = 3.6 cm]{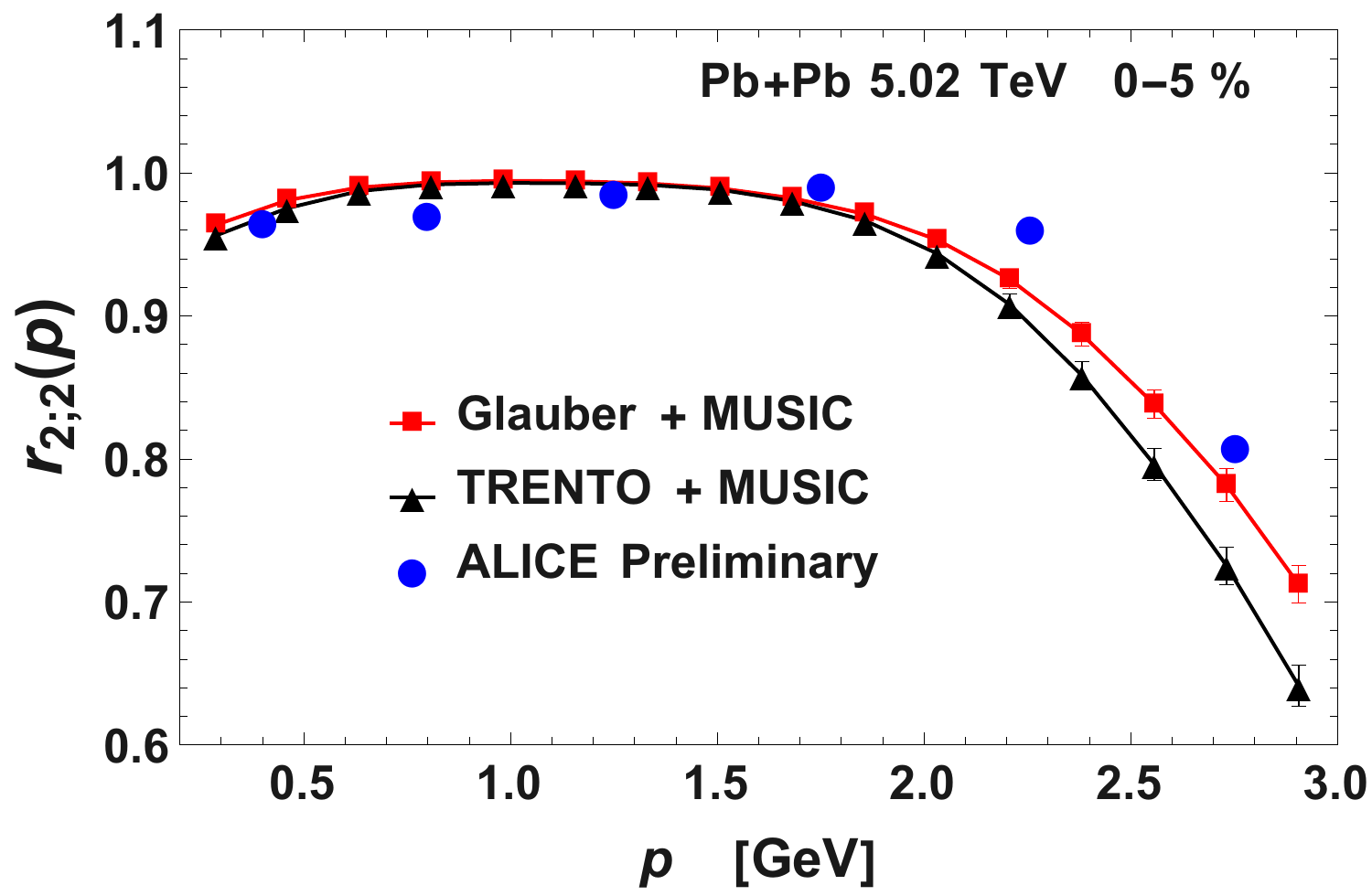}
	\end{subfigure}%
	\begin{subfigure}{.5\textwidth}
		\centering
		\includegraphics[height = 3.6  cm]{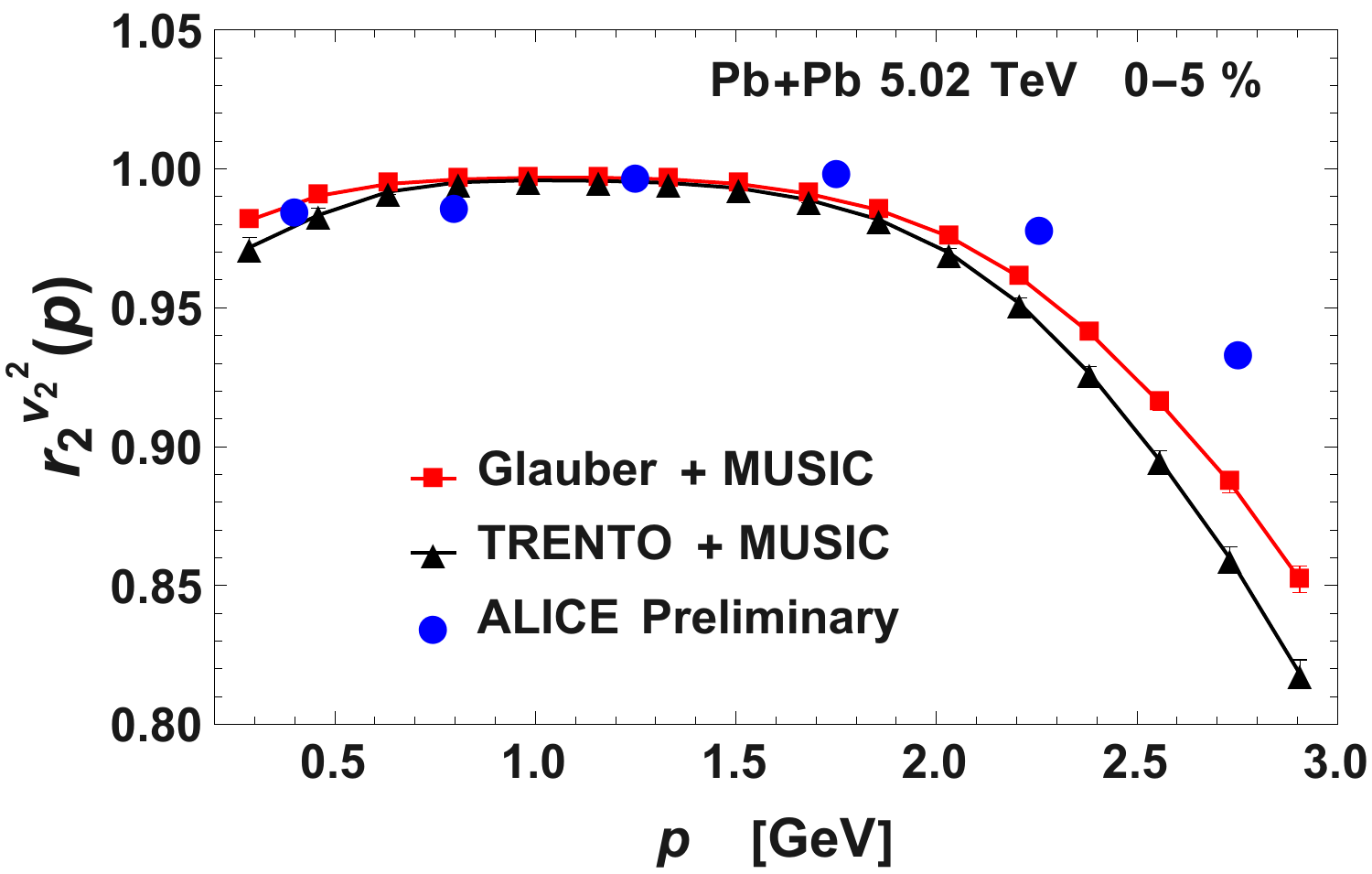}
	\end{subfigure}
	\caption{\footnotesize The factorization breaking coefficient between flow vectors squared $V_2(p)^2$ and $V_2^2$ (left) and for the flow magnitudes squared $v_2(p)^2$ and $v_2^2$ (right) for the elliptic flow. The red squares and black triangles denote the results obtained with the Glauber model and the TRENTO model initial conditions respectively. The experimental data of the ALICE Collaboration are represented using  blue dots. (Fig. from \cite{Bozek:2021mov})}
	\label{fig: r22}
\end{figure}

To ease this difficulty, an alternative approach proposed by the ALICE collaboration, would be to keep only one of the flow in a transverse momentum bin and the other flow averaged over the momentum range. Following this, we construct the factorization-breaking coefficient using the four-particle correlator for the flow vector squared and flow magnitude squared as\cite{NielsenIS2021}, 
\begin{equation}
	r_{n;2}(p)=\frac{\langle V_n^2 V_n^{\star }(p)^{2}\rangle}{\sqrt{\langle v_n^4 \rangle \langle v_n^4(p) \rangle}} \ \ and \ \ r_{n}^{v_n^2}(p)=\frac{\langle v_n^2 v_n^{ 2}(p)\rangle}{\sqrt{\langle v_n^4 \rangle \langle v_n^4(p) \rangle}} \ .
	\label{eq: flowsqfac}
\end{equation}  
We use the boost invariant version of the hydrodynamic model MUSIC \cite{Schenke:2010nt} for Pb+Pb collisions at $5.02$ TeV. For the initial state, two models are used; a two-component Glauber Monte Carlo model \cite{Bozek:2019wyr} and the TRENTO model \cite{Moreland:2014oya}.  We use a constant shear viscosity to entropy density ratio  $\eta/s=0.08$, unless otherwise specified.
In Fig. \ref{fig: r22}  are shown the factorization breaking coefficients between the flow vectors squared (left) and flow magnitudes squared (right) for the elliptic flow. For 0-5 \% centrality our model result corroborate the ALICE preliminary data. However, for semi-central collision (30-40 \%), our model results do not fully describe the data. Also one can see that the decorrelation for the magnitudes $r_2^{v_2^2}(p)$,  is approximately one half of the decorrelation for the vectors  $r_{2 ;2} ( p)$ in \ref{fig: r22},  which manifest the fact that the total flow vector decorrelation involves equal contribution from the flow magnitude and the flow angle decorrelation; $[1-r_{n;2}(p)]\simeq 2 [1-r_n^{v_n^2}(p)]$.

\begin{figure}
	\centering
		\centering
		\includegraphics[height = 3.6  cm]{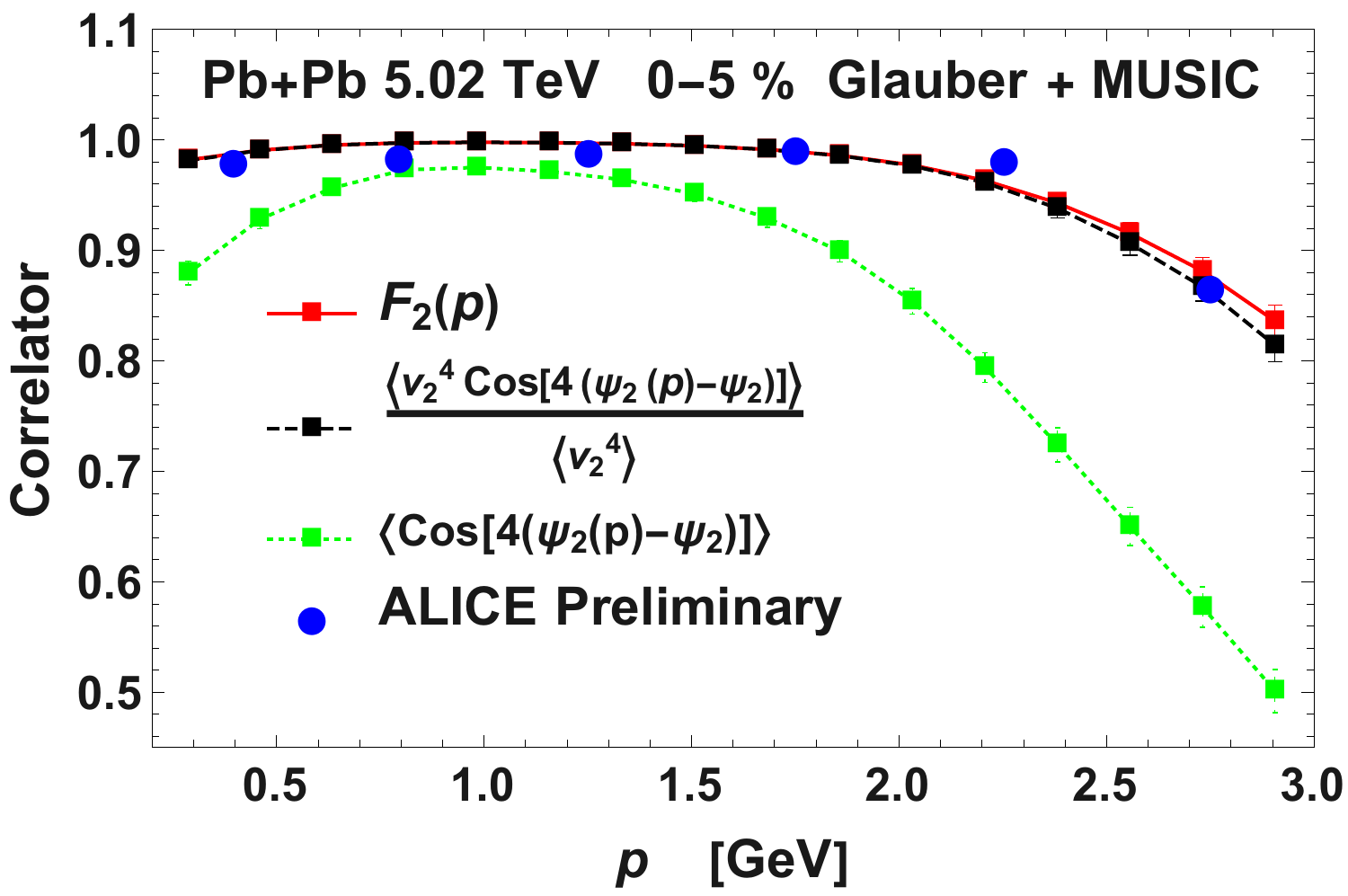}
	\caption{\footnotesize The flow angle correlation estimated as the ratio of the flow vector and the flow magnitude factorization breaking coefficients  (Eq. \ref{eq:ang}). The symbols with the dashed lines denote the angle correlations calculated directly from the simulated events (Eq. \ref{eq: wtdang}). The green squares represent the simple average of the decorrelation angle (Eq. \ref{eq:simav}).}
	\label{fig:angfac}
\end{figure}
 Once we have the flow vector and flow magnitude factorization-breaking coefficients separately, we can estimate the flow angle decorrelation by constructing the correlator for flow angle, taking the ratio of the four-particle flow vector correlator and flow magnitude correlator
Eq. \ref{eq: flowsqfac},
\begin{equation}
	F_{n}(p)=\frac{\langle V_n^2 V_n^{\star }(p)^{2}\rangle}
	{\langle v_n^2 v_n^{ 2}(p)\rangle} \ .
	\label{eq:ang}
\end{equation}
The above correlator  is an experimental measure of
the flow angle decorrelation:
\begin{equation}
	F_{n}(p)=\frac{\langle  v_n^2 v_n(p)^2 \cos\left[ 2 n  \left( \Psi_n(p) -\Psi_n \right)\right] \rangle}{\langle v_n^2 v_n(p)^2 \rangle} \ .
	\label{eq:angexp}
\end{equation}
However, the actual angle decorrelation could be defined as \cite{Bozek:2018nne},
\begin{equation}
\frac{\langle v_n^4 \cos\left[ 2n \left( \Psi_n(p)-\Psi_n\right) \right]\rangle}{\langle v_n^4\rangle} \ .
\label{eq: wtdang}
\end{equation}
The results presented in Fig. \ref{fig:angfac} indicate
that the two formulas give similar results, suggesting that the experimental measure \ref{eq:ang} could be used to estimate the weighted flow angle decorrelation. Our model results describe the experimental angle decorrelation. On the other hand, the simple average of the decorrelation angle
\begin{equation}
  \langle \cos\left[ 2n \left( \Psi_n(p)-\Psi_n\right) \right]\rangle
\label{eq:simav}
\end{equation}
is very different and cannot be estimated directly from the data.

\section{Mixed flow correlations}

Similar correlations coefficients can be constructed between  mixed flow harmonics,
 as a measure of non-linear coupling between the flow harmonics. Such correlation
coefficients are sensitive to the  non-linear response of the hot dense medium \cite{Bhalerao:2013ina}. For example, one can construct the correlation coefficients between
$V_2^2$ and $V_4 (p)$ or between $V_2V_3$ and $V_5 ( p)$. However, to extract the angle decorrelation between these mixed flow, one needs to construct the correlation coefficient (or factorization-breaking coefficients) between the squares of the harmonics used. The mixed flow vector-vector, magnitude-magnitude and angle decorrelation are given by,
\begin{equation}
\frac{\langle V_2^4  V_4^\star(p)^2\rangle }{\sqrt{{\langle v_2^8  \rangle}{\langle v_4^4(p)  \rangle}}}  \ , \frac{\langle v_2^4  v_4(p)^2\rangle }{\sqrt{{\langle v_2^8  \rangle}{\langle v_4^4(p)  \rangle}}} \ \ \rm{and} \ \ \frac{\langle V_2^4  V_4^\star(p)^2\rangle }{\langle v_2^4  v_4(p)^2\rangle } \ .
\label{eq:ang422}
\end{equation}
\begin{figure}
	\centering
	\begin{subfigure}{.5\textwidth}
		\centering
		\includegraphics[height = 3.6 cm]{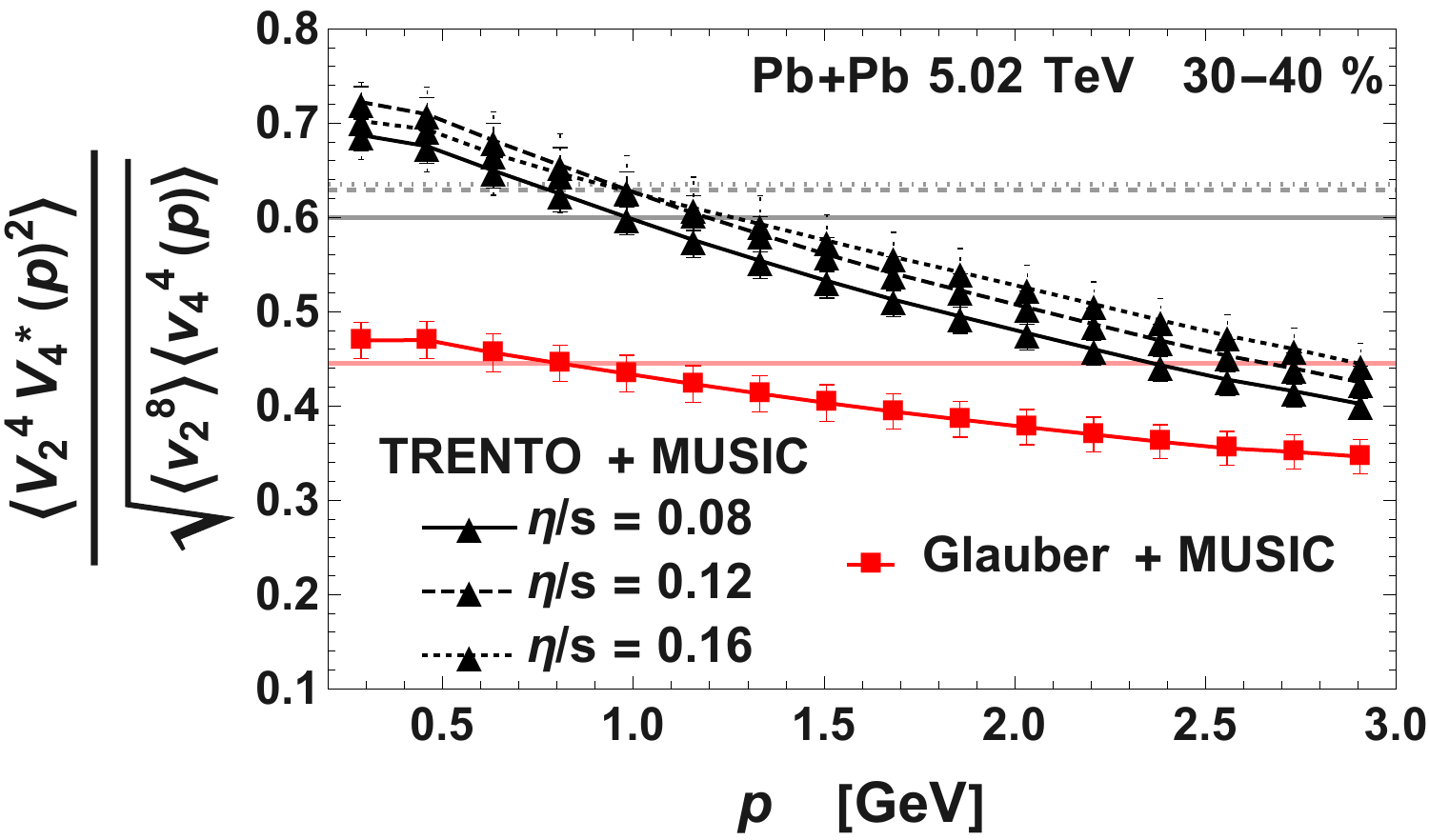}
	\end{subfigure}%
	\begin{subfigure}{.5\textwidth}
		\centering
		\includegraphics[height = 3.6  cm]{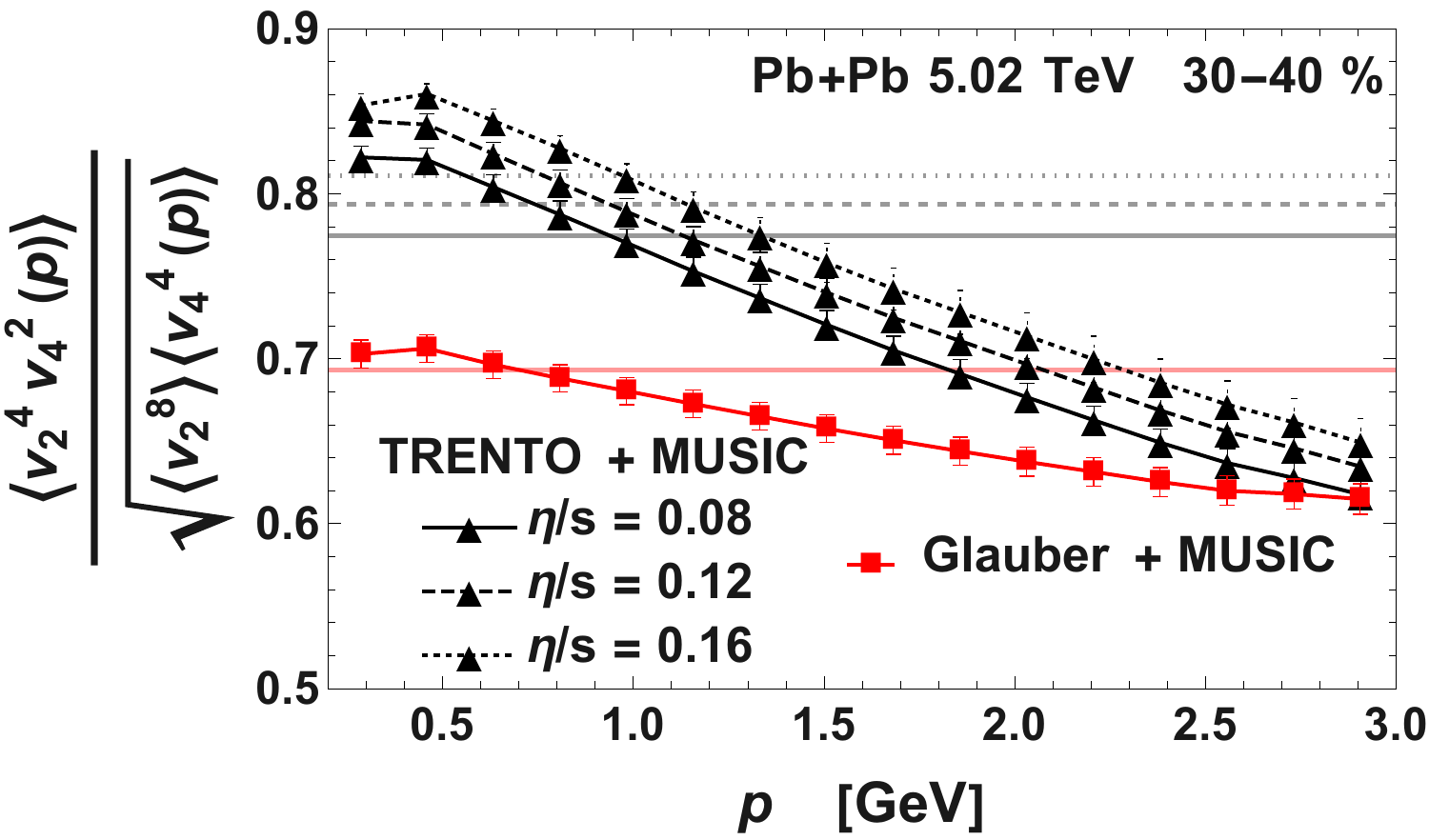}
	\end{subfigure}
	\caption{\footnotesize The flow vector (left) and flow magnitude correlation (right)  between  $V_2^4$ and $V_4(p)^2$. For the TRENTO model initial conditions, the solid, dashed and dotted lines represent results with $\eta/s=0.08,\ 0.12, \ 0.16$ respectively. The horizontal lines represent the corresponding correlation coefficients for momentum averaged flow vectors $V_2^4$ and $V_4^2$. (Fig. from \cite{Bozek:2021mov})}
	\label{fig:mixfac}
\end{figure}
\begin{figure}
	\centering
		\centering
		\includegraphics[height = 4.5  cm]{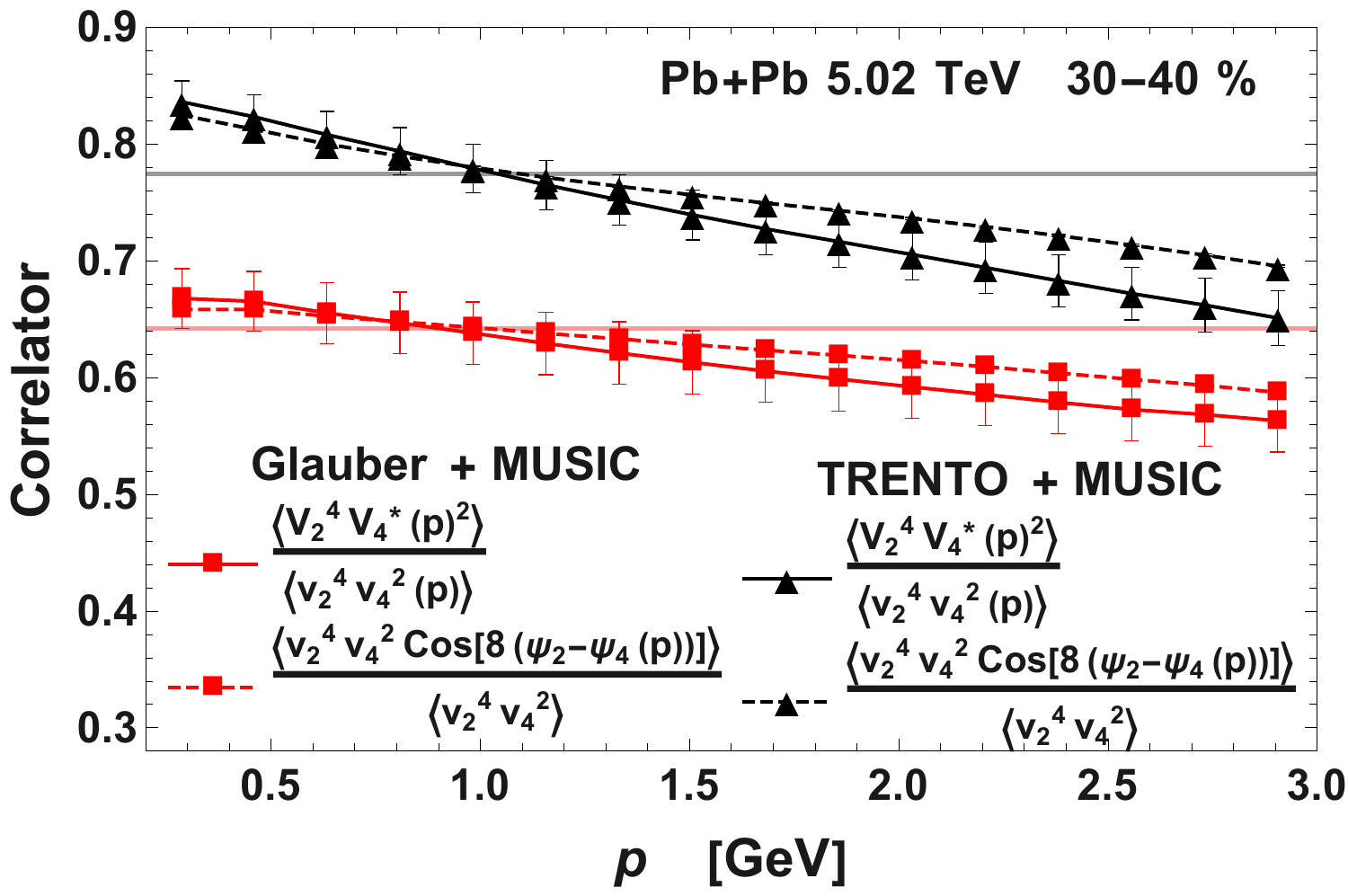}
	\caption{\footnotesize The flow angle correlation  between  $V_2^4$ and $V_4(p)^2$. (Fig. from \cite{Bozek:2021mov})}
	\label{fig:mixang}
\end{figure}


In Figs. \ref{fig:mixfac} and \ref{fig:mixang}, are shown the results predicted in the model for the mixed flow factorization-breaking coefficients between $V_2^4$ and $V_4^2$ for flow vector, magnitude and angle decorrelations.  Again, we notice that  the flow magnitude decorrelation is approximately one half of the
flow vector decorrelation. The horizontal lines denote the baseline for the correlations between the momentum averaged flow harmonics.

\section{Summary and outlook}
 We study the factorization-breaking coefficients between the harmonic flow coefficient in a fixed transverse momentum bin and the average flow harmonic. Such factorization coefficients show significant decorrelations which is a direct consequence of initial state fluctuation. The separate measurement of the flow vector and magnitude decorrelation lead to the extraction of flow angle decorrelation, where the magnitude and the flow angle decorrelations constitute approximately equal parts of the total flow vector decorrelation. Our model results well reproduce the ALICE preliminary data in central collisions, but deviate in case of  semi-central collisions. We also present predictions for the momentum dependent mixed flow correlations, which measure the non-linear response of the hydro response and put additional constrain on the models. Our results could be verified in experiments.  
\section*{Acknowledgments}
This research was partly supported by the  Polish National Science Centre grant: 2019/35/O/ST2/00357.
	
\bibliography{hydr.bib}	

\end{document}